# Leveraging tropical reef, bird and unrelated sounds for superior transfer learning in marine bioacoustics

Ben Williams[abc], Bart van Merriënboer[b], Vincent Dumoulin[b], Jenny Hamer[b], Eleni Triantafillou[b], Abram B. Fleishman[d], Matthew McKown[d], Jill E. Munger[de], Aaron N. Rice[fg], Ashlee Lillis[h], Clemency E. White[ij], Catherine A. D. Hobbs[ij], Tries B. Razak[k], Kate E. Jones[a], Tom Denton[l]

[a]University College London, UK. [b]Google DeepMind, London, UK. [c]Zoological Society of London, UK.[d]Conservation Metrics Inc., Santa Cruz, USA. [e]University of New Hampshire, USA. [f]K. Lisa Yang Center for Conservation Bioacoustics, Cornell Lab of Ornithology, Cornell University. [g]Department of Public and Ecosystem Health, Cornell University. [h]Sound Ocean Science, South Africa. [i]University of Exeter, UK. [j]University of Bristol, UK. [k]IPB University, Indonesia. [l]Google Research, San Francisco, USA.

## Abstract

Machine learning has the potential to revolutionize passive acoustic monitoring (PAM) for ecological assessments. However, high annotation and compute costs limit the field's efficacy. Generalizable pretrained networks can overcome these costs, but high-quality pretraining requires vast annotated libraries, limiting its current applicability primarily to bird taxa. Here, we identify the optimum pretraining strategy for a data-deficient domain using coral reef bioacoustics. We assemble ReefSet, a large annotated library of reef sounds, though modest compared to bird libraries at 2% of the sample count. Through testing few-shot transfer learning performance, we observe that pretraining on bird audio provides notably superior generalizability compared to pretraining on ReefSet or unrelated audio alone. However, our key findings show that cross-domain mixing which leverages bird, reef and unrelated audio during pretraining maximizes reef generalizability. SurfPerch, our pretrained network, provides a strong foundation for automated analysis of marine PAM data with minimal annotation and compute costs.

## Intro

Advanced monitoring tools are key to tackling the biodiversity crisis (Pimm et al., 2015). Passive acoustic monitoring (PAM) represents a powerful medium through which to gather data for ecological assessment and monitoring purposes (Gibb et al., 2020; Ross et al., 2023). Low-cost autonomous recording units are now widely available, enabling the collection of vast quantities of PAM data (Hill et al., 2018; Lamont et al., 2022; Shonfield et al., 2017). This has resulted in data collection scaling beyond the analytical capacity of human annotators (Gibb et al., 2020). Effective automated analysis is therefore required to maximize the potential of these data. Machine learning (ML) has emerged as a powerful tool with the potential to meet this demand, with state of the art approaches typically leveraging deep neural networks (Stowell, 2022). The current paradigm for ML-accelerated PAM analysis employs supervised learning techniques to train classifiers which can detect target signals. These supervised learning techniques are typically used to develop species-level detectors or identify anthropogenic activities (Gibb et al., 2020; Stowell, 2022).

A key drawback of traditional supervised approaches is their reliance on large annotated libraries of validated target sounds, typically requiring hundreds of examples per class to train an accurate classifier. These libraries are costly and time-consuming to assemble due to their reliance on human annotators (Kholghi et al., 2018). Additionally, classifiers often generalize poorly to "out of-distribution" data, where new data differs significantly from the initial training set (e.g., new field sites, different microphones). To address these issues, sophisticated efforts have sought to develop broad multi-species classifiers. These networks are trained on large and diverse libraries of recordings. Well-established pretrained bioacoustic networks have been primarily trained on avian taxa, where large open-source annotated libraries are available (e.g., Xeno-canto, Macaulay Library)(Ghani et al., 2023; Kahl et al., 2021). These pretrained networks can sometimes be used off the shelf to classify sounds new recordings. However, their use in this manner is restricted to only the classes present in their training set. Furthermore, these pretrained classifiers still underperform on datasets which are out-of-distribution from their training data and on classes which are under-represented in their training data (Stowell, 2022; Pérez-Granados, 2023).

When a pretrained network cannot directly classify novel signals, transfer learning represents an effective alternative. Here, samples from new target classes are passed through the pretrained network, and outputs from an intermediate layer are used to produce a feature embedding representation of the samples. These fixed embeddings can then be used to train a lightweight machine learning classifier (Ghani., et al 2023; White et al., 2023). This removes the costly training phase by leveraging knowledge from the network's embedding space learned during pretraining. Additionally, strong pretrained models facilitate few-shot transfer learning, where only a small number of training examples are needed to produce a highly accurate classifier, significantly reducing annotation costs (Nolasco et al., 2022). Emerging research shows networks pretrained on unrelated terrestrial bioacoustic data transfer well to similar bioacoustic domains, enabling few-shot learning (Ghani et al., 2023). However, the ability of pretrained bioacoustic networks to transfer to highly novel domains, such as aquatic environments, is largely untested (Williams et al., 2024). Substantial domain shifts may require the development of novel pretrained networks to achieve accurate few-shot transfer learning. The optimal pretraining strategies to produce these networks remain unknown, which is further compounded by the sparsity of annotated libraries for novel domains.

Coral reef ecosystems host some of the highest documented bioacoustic diversity in the ocean (Kaplan et al., 2018; McWilliam et al., 2018), yet ML-accelerated PAM analysis is significantly underdeveloped for these habitats. Coral reefs host >25% of marine biodiversity and >375M people are directly reliant on the ecosystem services these habitats provide (Hoegh-Guldberg et al., 2019; Knowlton et al., 2010). They are also among the most threatened habitats globally, with >50% of the world's reefs lost over the last 70 years and a projected loss of 90% of those remaining by 2050 (Eddy et al., 2021; IPCC 2019). The soundscapes of these habitats have been found to contain information relevant to key ecosystem attributes such as coral cover and fish community diversity, as well as representing a key ecosystem function which drives larval recruitment (Kaplan et al., 2018; Pysanczyn et al., 2023). PAM is therefore emerging as a promising tool to monitor these habitats (Mooney et al., 2020), but there is a sparsity of relevant annotated data. As is common for underwater soundscapes, the majority of biological sounds on coral reefs remain un-documented and for a significant portion of those that have been recorded, the taxonomic source of origin has not been validated (Parsons et al., 2022; Rountree et al., 2019). Automated analysis of reef PAM data is therefore highly underdeveloped, with only a very limited number of studies having used ML-accelerated analysis on PAM data from these habitats (Lin et al., 2018; Ozanich et al., 2021; Williams et al., 2022). Consequently, coral reef habitats represent an exemplary candidate for exploring novel few-shot transfer learning bioacoustic frameworks, with advances in this field having the potential to help address real-world conservation challenges.

In this study, we aim to identify the optimum pretraining strategy to produce a network which supports accurate few-shot transfer learning for a novel bioacoustic domain. Such a network should facilitate analysis of PAM data with minimal computational and annotation costs. We selected the coral reef domain due to the threatened status and high acoustic diversity of these ecosystems, with the potential to provide a strong foundation for transferring to other aquatic habitats. To achieve this, we first assembled ReefSet, a meta-dataset that consists of 57K annotated 1.88 sec recordings across 37 classes from 16 unique datasets. Although only 1.99% the size of comparable bird libraries at the time of writing (Xeno-canto, 2023), ReefSet represents the largest published dataset of annotated reef recordings to date. Data from aquatic environments represents a large domain shift from commonly used pretrained networks which are primarily trained on bird or more general audio data. Our first experiment therefore set out to determine how well existing pretrained networks perform at few-shot transfer learning on ReefSet. Next, we tested whether pretraining on this significantly smaller but in-domain dataset improves upon the performance of existing pretrained networks. To assess the generalizability of this strategy across the reef domain, we employed a rigorous evaluation of generalization protocol which consisted of training on 15 constituent datasets within ReefSet whilst holding one out for testing, which was repeated in every combination one by one. We then, tested whether leveraging cross-domain mixing during pretraining can improve performance, following the same rigorous evaluation protocol. Finally, we tested whether the output from cross-domain mixing optimised for the coral reef domain impacts generalizability to unrelated domains.

## Dataset

To maximize the generalizability of our approach, we compiled a diverse meta-dataset of labeled coral reef bioacoustic recordings from 16 individual datasets across 12 countries (Fig. 1; Supp. 1. Table S1), hereon referred to as “ReefSet”. Each individual dataset was originally collected and labeled for different purposes using a variety of sampling strategies and hydrophone models (Supp. 1. Table S2). During annotation of each dataset, longer recording periods were cut into samples of shorter windows (1.88 sec) and labeled by human annotators using aural and visual inspection of each sample's spectrogram. While many classes were of a known origin, others were unknown but typically presumed to originate from fish. All labeled samples were then re-sampled to 16 kHz and written out as a separate waveform audio file.

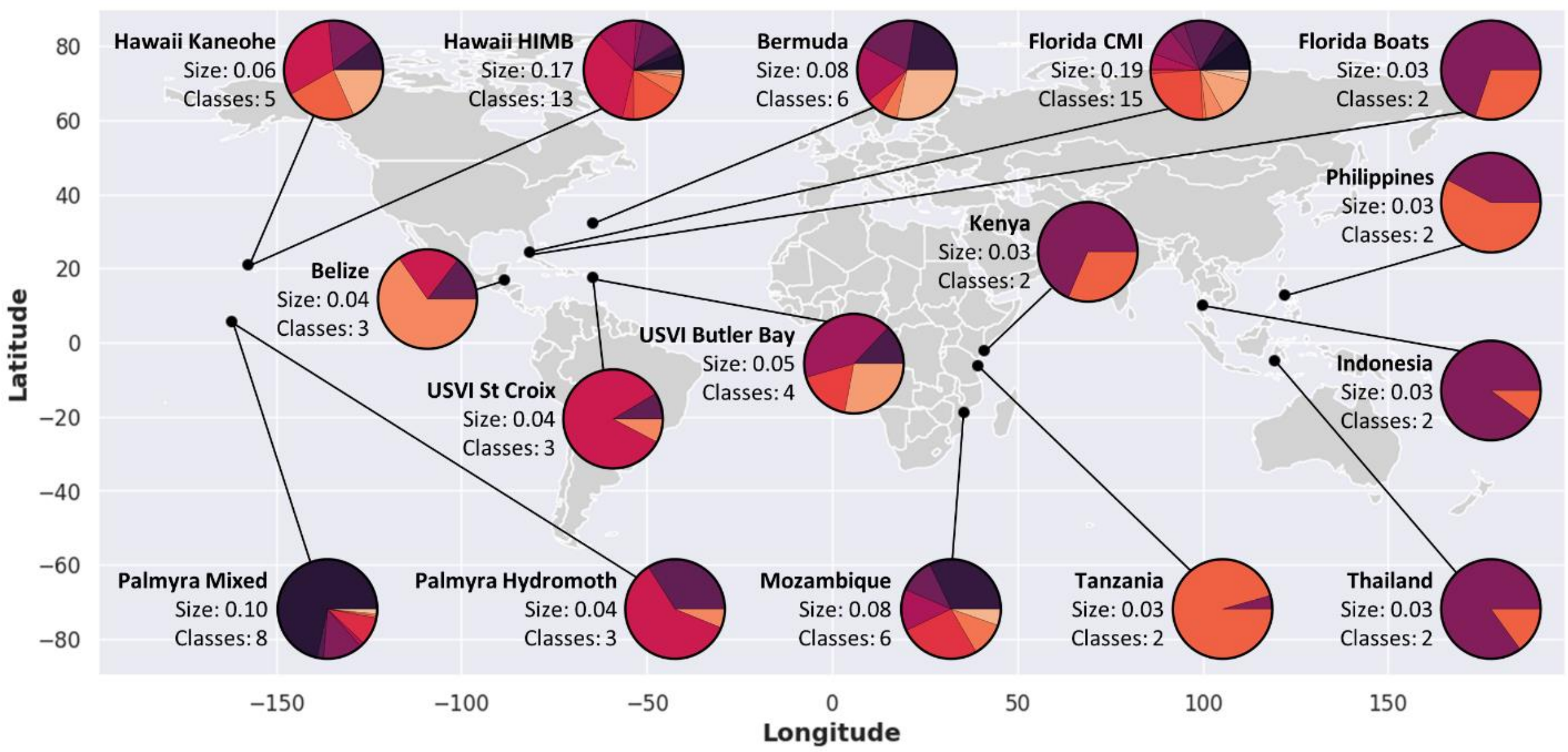

**Fig. 1.** Datasets used to assemble ReefSet. Size indicates the relative size of each dataset to ReefSet, summing to one. Classes indicates the number of unique labels within each dataset. Pie charts indicate the distribution of labels within the dataset.

To amalgamate class labels across datasets, each sample was first given a primary label: biophony, anthrophony, geophony or ambient. Here, the ambient label was used for negative samples in the Florida-boats, Kenyan and Indonesian datasets where the annotation strategy used a positive label class (motorboat, fish noise and bomb fishing respectively) alongside a strongly labeled negative class. A secondary label was then applied to all samples using existing labels from the datasets, with merging of labels under a common name where sounds matched across multiple datasets (e.g motorboats). Later, for evaluation, classifiers were trained on a maximum of 32 samples per class for each dataset, with a minimum of 10 samples from each class held out for testing. Therefore, classes with less than 42 samples in any given dataset were merged by only applying a primary label (biophony, geophony or anthrophony). Where the count of samples merged under the primary label class still did not total 42 or more samples in a given dataset, these samples were discarded. This yielded the final meta-dataset of 57,074 labeled samples, split across the four primary labels: biophony (79.20%), anthrophony (10.39%), geophony (0.09%) and ambient (10.32%), with 33 secondary labels (Supp. 1. Table S2).

## Results

### Pretrained bioacoustic networks outperform networks pretrained on general audio

Our first experiment was designed to test the generalizability of existing pretrained networks to the highly novel domain of coral reef bioacoustics. We identified four pretrained networks widely adopted for use in acoustic transfer learning (Table 1). All four networks employ a convolutional neural network architecture. The first two were VGGish (Hershey et al., 2017) and YAMNet (Google Research), both trained on general-purpose audio datasets. VGGish was trained on the YouTube-70M dataset, a dataset consisting of 20B weak multi-label samples across 31K classes. YAMNet was trained on AudioSet, a large ontology of 2.1M human labeled acoustic events across 521 classes gathered from YouTube (Gemmeke et al., 2017). The second two were BirdNET (Kahl et al., 2021) and Perch (Ghani et al., 2023) which were both primarily trained on bird recordings from the Xeno-Canto repository. Perch was trained on the full corpus of Xeno-Canto (XC) bird recordings split into 2.9m samples 5 sec in length, and was configured with hierarchical taxonomic output heads for species, genus, family, and order classes. BirdNET was trained on a smaller set of bird classes than Perch overall, but included bird samples from the Cornell Lab of Ornithology's Macaulay Library (Macaulay, 2023) alongside 101 additional classes such as human speech, dogs and amphibian species.

| Network | Training domain | Number of training classes | Input sample rate (kHz) | Input length (sec) | Embedding dimension | Parameter count | Real-time-factor inference speed (CPU) |
|---|---|---|---|---|---|---|---|
| VGGish | AudioSet (YouTube) | 31000 | 16 | 0.96 | 128 | 72.1M | 41.1 |
| YAMNet | AudioSet (YouTube) | 521 | 16 | 0.96 | 1024 | 4.7M | 86.04 |
| BirdNET v2.3 | Bioacoustic (primarily birds) | 3337 | 48 | 3 | 1024 | 10.4M | 260.68 |
| Perch v1.4 | Bioacoustic (birds) | 10932 | 32 | 5 | 1280 | 80.1M | 39.41 |

**Table 1.** Details of the four pretrained networks used to evaluate transfer learning performance on ReefSet. Real-time-factor inference speed reflects how many times faster than the audios real time duration each network is at processing audio on a CPU, further details are in Supp. 1.

To evaluate the transfer learning capabilities of the four pretrained networks, each network was configured with a final classification head which was fine-tuned whilst the rest of the weights were kept frozen. One by one, networks were configured and fine-tuned for all 16 datasets. This was repeated using 4, 8, 16 and 32 training examples per class. Mean AUC-ROC scores revealed the pretrained networks ranked consistently across all four training sample counts, in ascending order the mean and standard deviations of these were: BirdNET (0.908 ±0.09), Perch (0.881 ±0.11), YAMNet (0.834 ±0.05), VGGish (0.813 ±0.05) (Fig. 2). These results revealed the two networks pretrained primarily on the bird domain outperformed the two trained on the more general YouTube data. As expected, the mean AUC-ROC of all pretrained models improved and standard deviation of this declined as the number of training samples per class increased from 4 through to 32 (Fig. 2).

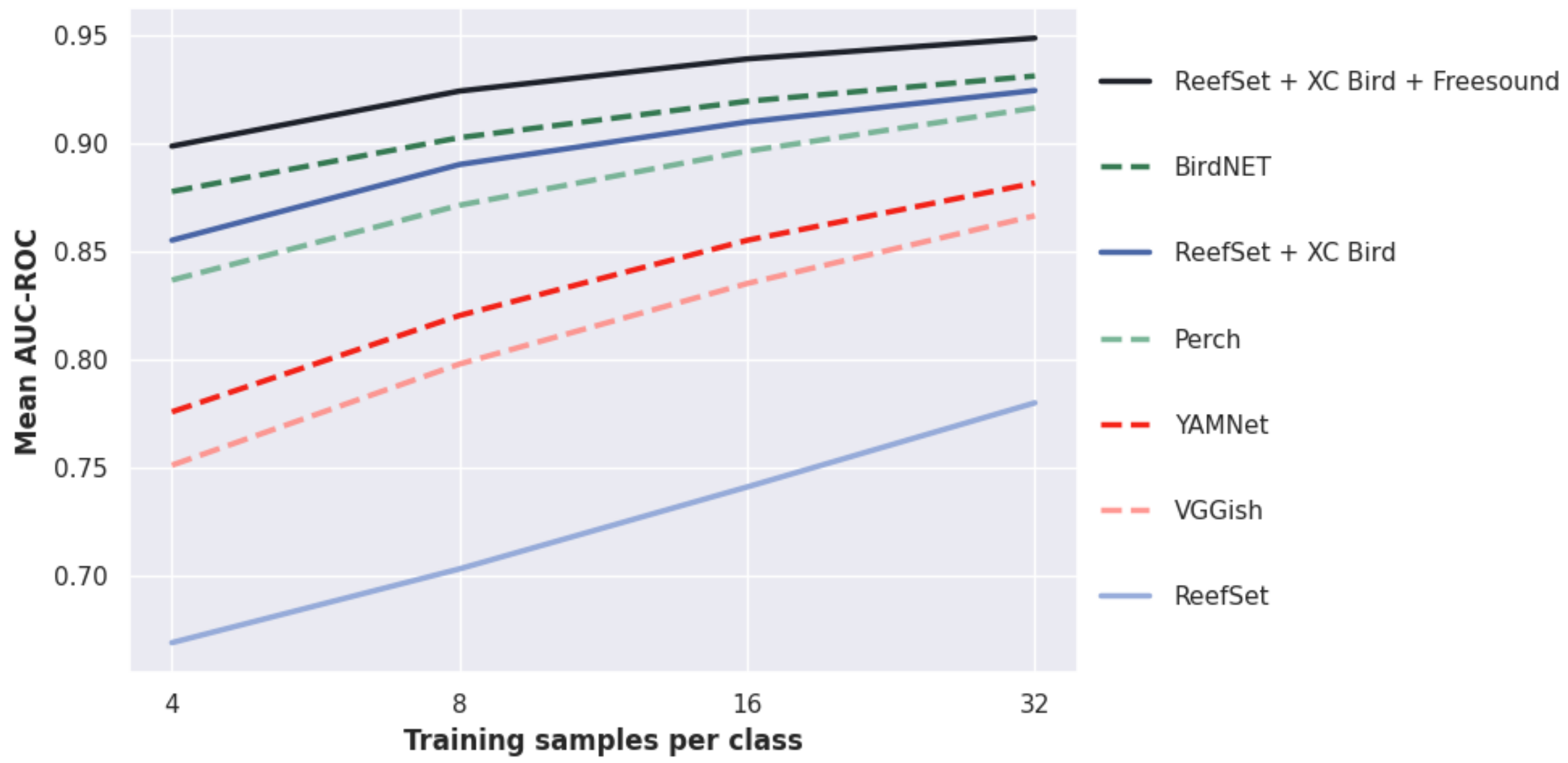

**Fig. 2.** Mean AUC-ROC scores reported by transfer learning evaluation for each model across all 16 datasets within ReefSet. Dashed lines represent existing pretrained networks, with names indicated in the legend. Thick lines represent our alternative pretraining strategies, with the data used for pretraining indicated in the legend.

Considering the constituent datasets within ReefSet on an individual basis, these presented a range of apparent difficulty (Fig. 3; Fig. S1). The datasets from Thailand, the Philippines and Indonesia, which only required binary classification between a one anthropogenic and one biophony class, generally presented easier challenges with mean AUC-ROC scores of 0.994 (±0.007), 0.960 (±0.010) and 0.935 (±0.056) respectively across all four pretrained networks and sample counts. More challenging datasets were those which required prediction of a small number of classes where samples were labeled with secondary biophony labels alongside samples labeled only with the primary biophony label class. These more challenging tasks included the Kenya, Belize and Tanzania datasets with mean AUC-ROC scores of 0.703 (±0.043), 0.791 (±0.065) and 0.812 (±0.037) respectively across all four pretrained networks and sample counts. Using just four training examples per class with BirdNET, the overall best performing pretrained network, the lowest and highest AUC-ROC scores were reported for the Kenya and Thailand datasets, with mean AUC-ROC scores of 0.746 (±0.062) and 0.996 (±0.006) respectively across the ten random seeds used for each.

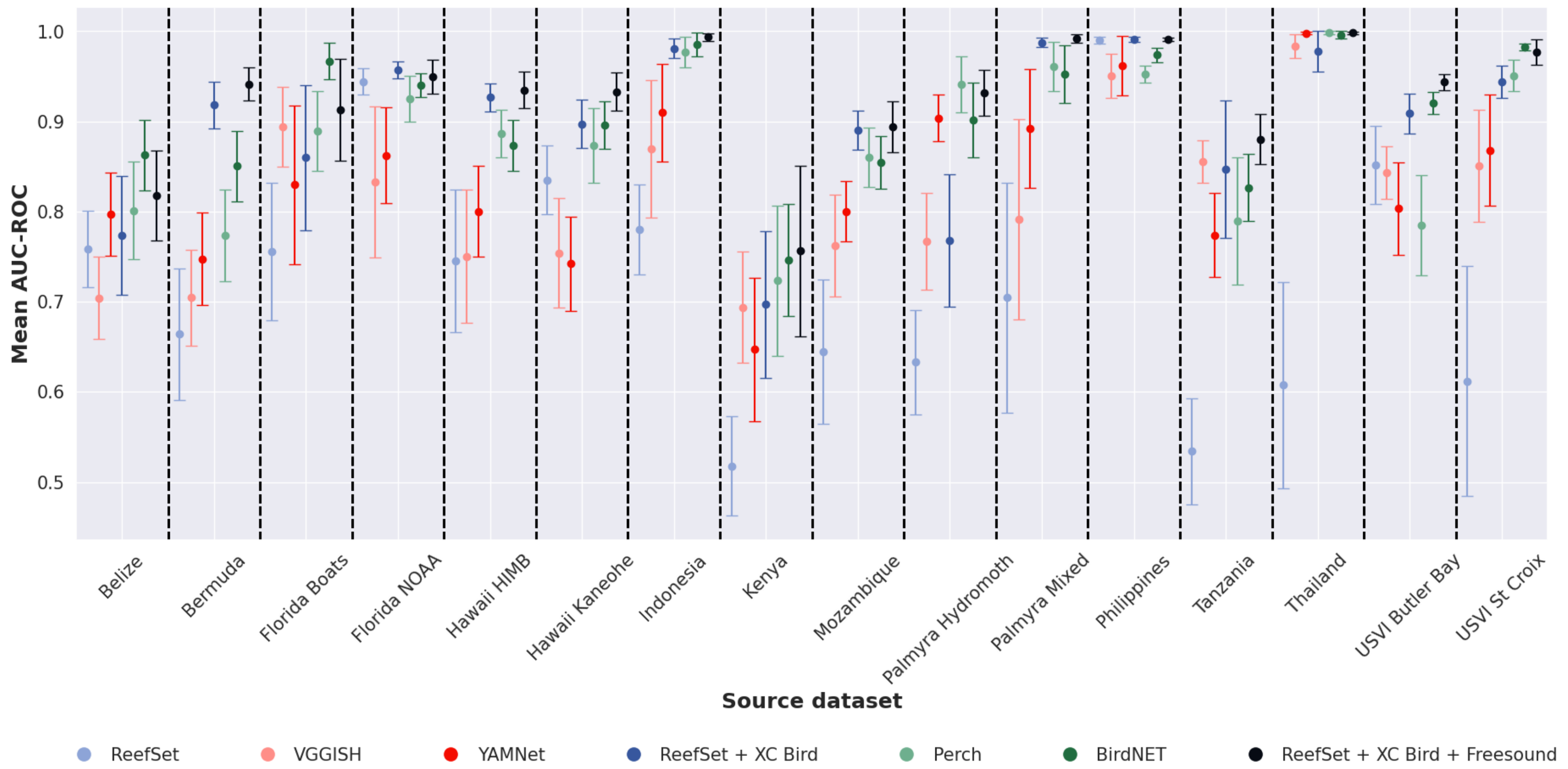

**Fig. 3.** Mean AUC-ROC scores reported by transfer learning evaluation on each dataset within ReefSet across all training samples per class counts used (4, 8, 16 and 32). Points represent the mean, lines represent standard deviation. Within each dataset bin, models are ordered by overall mean performance across all dataset and training sample counts, going from weakest (left) to strongest (right). In the legend, existing pretrained networks are indicated by name, whereas our alternative pretraining strategies are indicated by the datasets used during pretraining

### Existing pretrained networks outperform pretraining on limited in-domain data

Our second experiment revealed that few-shot transfer learning capabilities of a CNN pretrained on our highly in-domain but smaller ReefSet meta-dataset were considerably lower than that of all four existing pretrained networks. To rigorously test this, we used a dataset rotation for evaluation of generalization (DREG) approach which involved pretraining on 15 of the 16 datasets within ReefSet, whilst holding one out for evaluation, this was repeated one by one in all combinations. This strategy reported a mean AUC-ROC score of 0.724 (±0.05) across all four training sample counts, lower than any of the pretrained networks. BirdNET and VGGish, the highest and lowest performing pretrained networks respectively, had a 200.72% and 47.80% lower AUC-ROC error (area above the ROC) than the ReefSet only pretraining strategy.

### Cross-domain pretraining improves generalizability

Our third experiment revealed cross-domain mixing of the small in-domain ReefSet dataset with a large set of out-of-domain bioacoustic data can be leveraged during pretraining to provide considerable improvements. Given the performance of models pretrained using bioacoustic data identified during our prior experiments (Fig. 2), we tested mixing the full XC Bird catalog of 2.9M samples used to train Perch with the more modest ReefSet, approximately 1.99% of the size in total sample count (Fig. 4). This mixing provided a total of 10’165 target classes, less any exclusive to held-out reef data for DREG evaluation. Using the DREG pretraining and evaluation procedure once again, we observed a mean AUC-ROC score of 0.895 (±0.03) across all four training sample counts using this cross-domain pretraining strategy (Fig. 2). This represented a notable improvement over pretraining on the in-domain ReefSet alone, with a 163.90% reduction in AUC-ROC error. Importantly, this also achieved an improvement upon pretraining with the XC Bird dataset alone, represented by the pretrained Perch model, with a 12.32% lower AUC-ROC error. The only pretrained network which still outperformed this cross-domain pretraining strategy was BirdNET, with a 12.24% lower AUC-ROC error than our ReefSet and XC Bird cross-domain pretraining strategy.

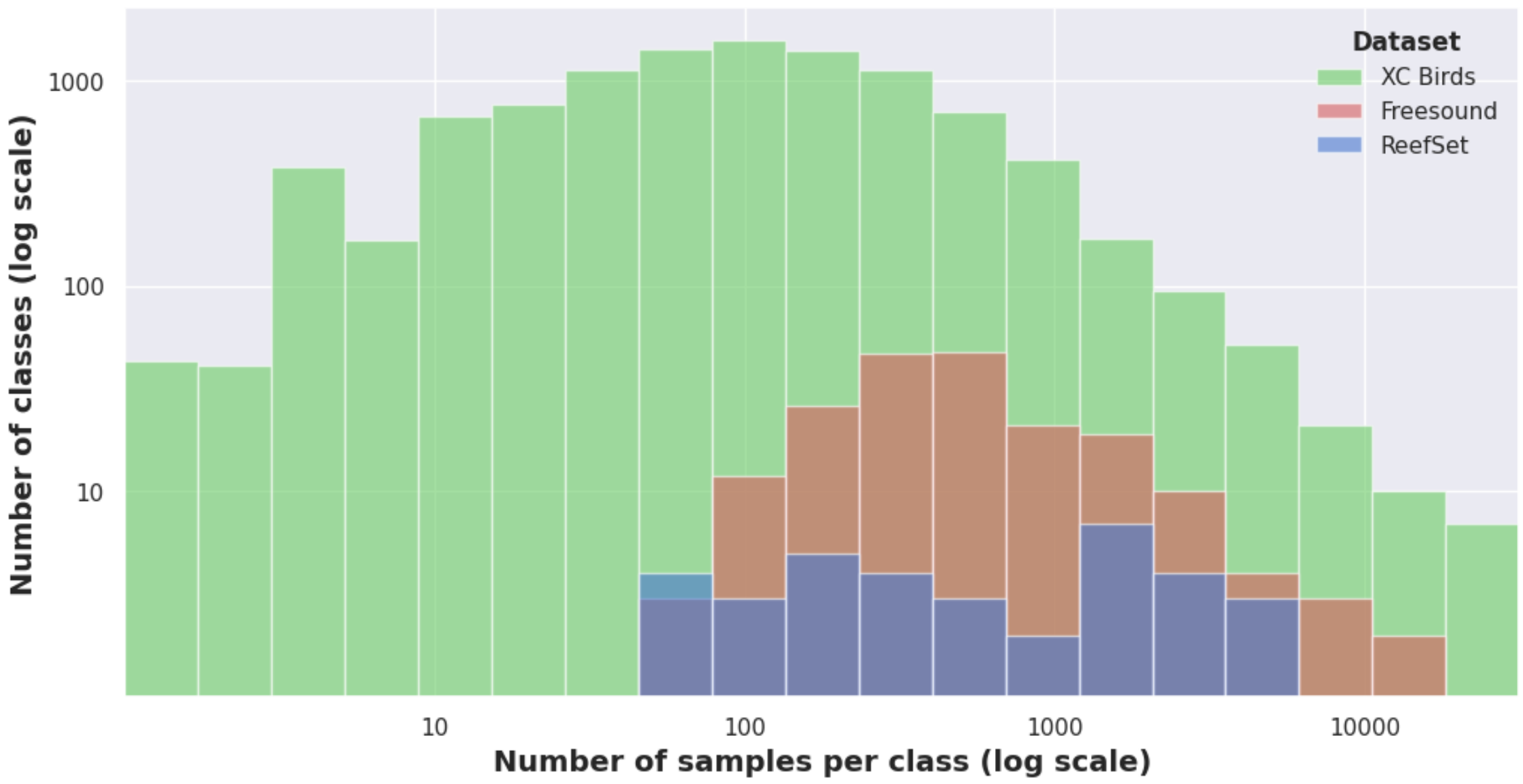

**Fig. 4.** Histogram of counts by class for the three datasets used for pretraining: XC Birds, Freesound and ReefSet. Bins are logarithmically spaced based on the range of counts in the XC Bird data, with bin edges determined by creating 20 equal intervals on a logarithmic scale between the minimum and maximum counts observed in the XC Bird dataset.

Our fourth experiment revealed that expanding the diversity of data used in cross-domain pretraining further enhanced few-shot transfer learning capabilities on the novel coral reef domain. Alongside ReefSet and the XC Bird dataset we mixed in Freesound Dataset 50K (Fonseca et al., 2022), a dataset based on the AudioSet ontology consisting of 108.2 hrs of annotated sound events across 200 classes (Fig. 4). Freesound is comprised of audio from a more general selection of sound events, comparable to the domain used to train VGGish and YAMNet, but with fully open source access. Using our DREG train and evaluation protocol, this triple-domain pretraining achieved the highest mean AUC-ROC scores of any strategy, with a mean AUC-ROC of 0.928 (±0.02) across all four training sample counts (Fig. 2). This represented a 31.26% reduction in error compared to cross-domain pretraining with the ReefSet and XC Bird datasets. Importantly, this strategy also outperformed BirdNET, the previously highest scoring network, with a 21.68% reduction in AUC-ROC error. Using just four training samples per class, this triple-domain pretraining strategy achieved a mean AUC-ROC of 0.900 (±0.02) across the 16 datasets.

Final trials using the triple-domain strategy revealed modifications to the bias, gain and smoothing parameters of the PCEN spectrogram (Supp. 1), alongside pretraining for 1m steps, further improved performance. Following the DREG pretraining and evaluation protocol, a mean AUC-ROC score 0.933 (±0.02) was reported using these adjustments, representing the strongest overall performance. This improvement corresponded to a 26.84% and 6.60% mean reduction in AUC-ROC error against BirdNET and our initial triple-domain pretraining trial respectively (Supp 1. Fig. S2). Finally, we produced SurfPerch, the open-source version of this model (Supp. 2), by pretraining with this triple-domain strategy including the full ReefSet meta-dataset, using all 16 source datasets.

**Targeted cross-domain pretraining does not improve generalizability to non-target bioacoustic domains**

Whilst this approach reports notable generalizability improvements to the reef domain, we observe this strategy negatively impacts performance on alternative bioacoustic domains which were not optimized for during pretraining (Fig. 5). Using the pretrained SurfPerch network, we mirrored the evaluation protocol used in Ghani et al. (2023) to assess the ability of Perch to generalize to novel datasets. These datasets originate from bird, frog, bat and marine mammal domains. Sample counts ranging from 4 to 256 training samples per class were used, with 10 repeats of each. We observed that Perch and BirdNET outperformed SurfPerch in all six of the novel domains. The lowest performance gap between SurfPerch and Perch, the overall best performing pretrained network at the novel challenges, was observed for the Godwit Calls and Watkins Marine Mammals datasets, with mean AUC-ROC scores 0.019 lower than Perch for both datasets across all training sample counts per class. The largest performance gap between SurfPerch and Perch was observed for the Yellowhammer dialect dataset, with a mean AUC-ROC score 0.084 lower than Perch across all training sample counts per class. However, SurfPerch did outperform both YAMNet and VGGish across all datasets. As the training samples per class counts increased, the performance gap between SurfPerch and Perch decreased for each, with a difference between mean AUC-ROC scores across all datasets of 0.067 for 4 training samples per class, and 0.012 for the maximum training sample count per class (32 or 256).

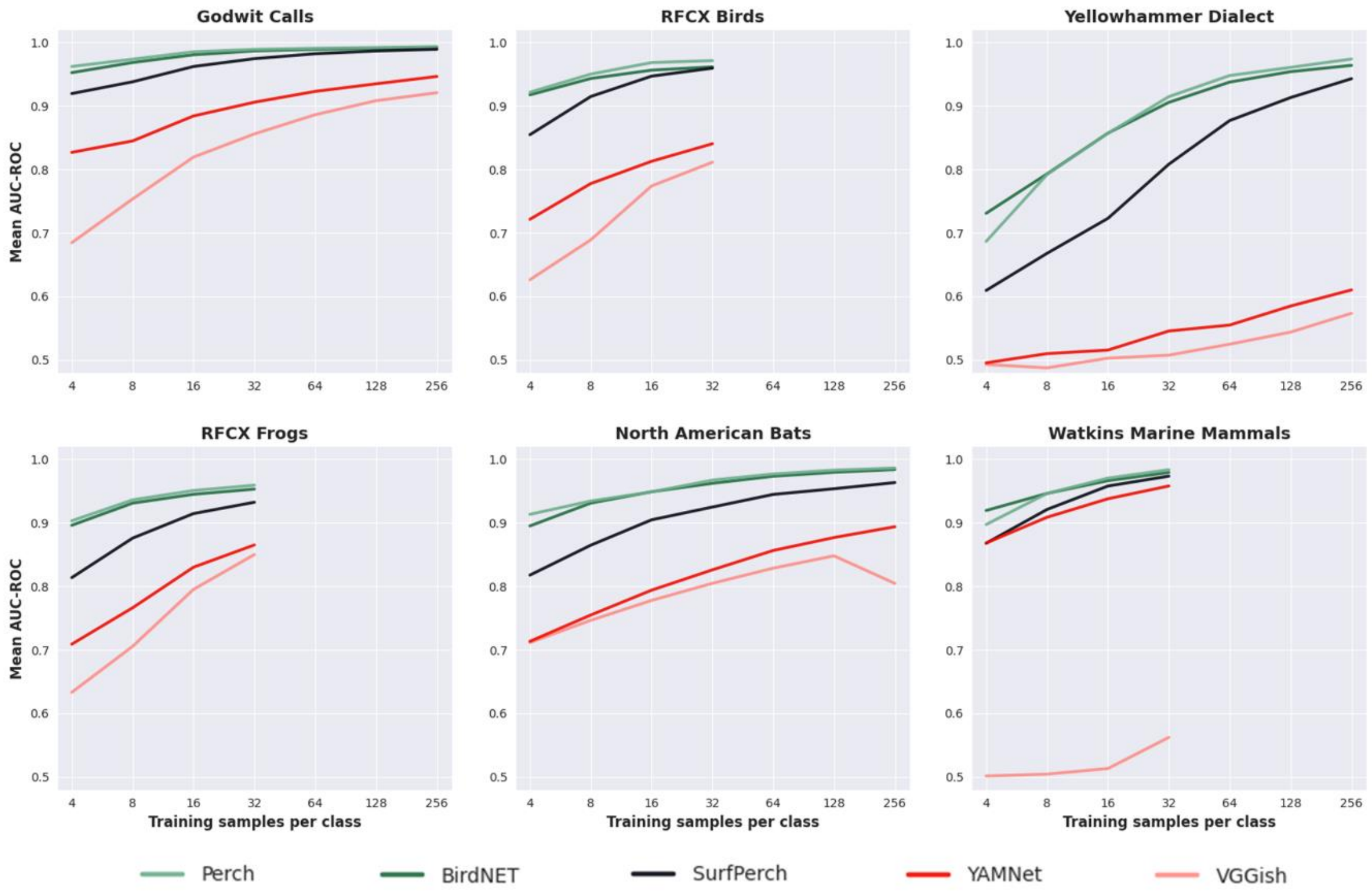

**Fig. 5.** Mean AUC-ROC scores reported from transfer learning evaluation of each model on six novel bioacoustic datasets. Each model and training sample per class count combination were repeated across ten random seeds.

## Discussion

Our findings show that leveraging multiple domains during pretraining provides superior transfer learning capabilities to the data-deficient domain of marine bioacoustics. We began by testing the transfer learning capabilities of existing pretrained networks on marine bioacoustic data. We found networks pretrained on data from bioacoustic domains outperformed those pretrained with more general audio data from unrelated domains. Next, we show these existing pretrained networks outperformed pretraining with our highly in-domain, but smaller, ReefSet meta-dataset. We then found that cross-domain mixing using the larger out-of-domain XC Bird dataset with the smaller in-domain ReefSet improved upon transfer learning capabilities of the previous strategies. Finally, we reveal that mixing together all three domains during pretraining provides the strongest performance. However, we observe that cross-domain mixing did not improve performance on unrelated bioacoustic domains which were not optimized for during pretraining. These findings have implications for the field of bioacoustics and other acoustic domains that do not currently have adequate pretrained networks.

The product of our optimum strategy, SurfPerch, supports accurate few-shot learning, achieving a mean AUC-ROC of 0.902 (±0.09) using just four training samples per class. Furthermore, this was evaluated by only fine-tuning the final layer of the network during transfer learning. This combined few-shot transfer learning protocol therefore significantly reduces both the annotation and computational costs required to build accurate classifiers, enabling end-users to perform this on standard personal computing devices using a minimal set of annotations. Supplementary 2 offers an interactive demonstration on how to implement this approach on new data from a web browser, including the incorporation of an agile modeling protocol which can be used to further boost classifier performance by identifying the most relevant samples for annotation (Hamer et al., 2023).

The results we present here provide key learnings on pretraining for bioacoustics. The performance of existing pretrained networks supports similar work in Ghani et al. (2023) which reported networks pretrained on large and diverse bird bioacoustic datasets generalize better to other bioacoustic domains than those pretrained on more general audio. The reasons behind this remain an open research question, this could be due to common properties between signals across bioacoustic domains, or, the high innate acoustic complexity and variety of bird vocalizations compared to AudioSet. An additional consideration which may contribute to the performance gap is that both bioacoustic networks were pretrained on a far larger diversity of classes compared to YAMNet, whereas VGGish was trained on a yet larger diversity of classes but these were weakly labeled (Table 1). Given that higher class diversity using strong labels during pretraining has been reported as a key driver behind few-shot performance (Luo et al., 2023), these label setups may well have contributed to the performance gap. Of the two bioacoustic networks, the stronger performance of BirdNET contradicts the label count consideration given its lower class count. However, the performance of BirdNET further supports the hypothesis that incorporating cross-domain mixing during pretraining improves generalizability to domains with only small annotated libraries as BirdNET (v2.3) was trained on a limited set of labels beyond bird samples, including: invertebrate, amphibian, mammal, human and anthropogenic sound classes.

Our findings present guidance for future users aiming to leverage deep neural networks for bioacoustics. Pretraining experiments such as those presented here are inherently computationally expensive. The experiments outlined here total to the training of 139 networks, on average requiring ~20 hrs on a TPUv3 pod, the equivalent of 26 '668 USD using Google Cloud spot instances at the time of writing. Alternatively, fine tuning a model is arbitrary, taking <1 min with 128 training samples using a standard personal laptop. Bioacousticians with novel datasets will therefore benefit most from identifying an existing pretrained network most relevant to their domain, or testing a suite of existing networks as presented in our first experiment. Here, we show SurfPerch presents the strongest option for coral reefs and likely related aquatic domains, whereas Perch, trained exclusively on bird data, presents a better option for unrelated bioacoustic domains. Interestingly, Perch still out performed SurfPerch on the Watkins marine mammal dataset. Many of the sounds in the Watkins dataset were in fact terrestrial marine mammal vocalizations which may partially explain this (Sayigh et al., 2016). Additionally, marine mammal calls typically consist of multiple high frequency phenomes (Erbe et al., 2017) potentially making these more comparable to the bird domain than the single phenome sounds characteristic of most fish vocalizations (e.g grunts, pops).

Where an adequate pretrained network is not available and domain specific data is sparse, we show cross-domain pretraining presents a valuable strategy to develop a suitable network for the target domain. Future work may be able to improve upon the strategies tested here. Increasing the volume of high quality in-domain training data is typically the most valuable tool for improving model performance (Halevy et al., 2009). Relevant sound libraries to the coral reef domain with potential for growth include the FishSounds platform (Looby et al., 2023) and the proposed Global Library of Underwater Biological Sounds (Parsons et al., 2022). Furthermore, growing open-source sound event libraries are available, meaning additional bioacoustic and unrelated acoustic domains could be integrated, (Cañas et al., 2023; Humphrey et al., 2018; Mac Aodha et al., 2018). More broadly, mixing data from a diverse set of domains and optimizing for a range of these during evaluation, rather than the single domain we targeted here, may present a route to developing improved foundational bioacoustic and acoustic models for wider contexts.

Beyond enhancing the data used for training, future work could also explore methodological improvements. Marine PAM data is inherently noisy, with recordings typically comprised of multiple sources from biophony, geophony and anthropogenic components (Mooney et al., 2020). Implementing an unsupervised source separation model presents a proven tool for improving classification in noisy bioacoustic datasets, through disentangling individual signals from the broader soundscape (Denton et al., 2022; Lin & Tsao, 2019). Elsewhere, self supervised learning (SSL) presents a tool which can be used to learn informative features from unlabeled data, enabling it to exploit vast un-annotated datasets (Moummad et al., 2023). Future work could benchmark pretraining on larger in-domain datasets with SSL against cross-domain pretraining of supervised classifiers. Other changes during the transfer learning component could boost performance. Firstly, alternative lightweight classifiers (e.g two layer models, random forests) could be tested. Augmentations are another proven way to improve classification accuracy with limited training data (Stowell, 2022). Better still, agile modeling, alternatively known as active learning, can be integrated into the pipeline using a human in the loop to identify the most informative training samples (Hamer et al., 2023; Stretcu et al., 2023).

In conclusion, we leveraged cross-domain pretraining to develop a powerful tool that supports automated analysis of marine PAM recordings with low annotation and computational costs for the end-user. Our findings offer insights into replicating this for other acoustic domains where existing pretrained networks are inadequate. Combining efficient machine learning analysis such as ours with the vast scales at which PAM data can be collected has a significant potential to boost our understanding and monitoring capacity of global biodiversity. We anticipate these technologies will facilitate the expansion of scientific frontiers towards new applications and challenges so far unrealized.

## Methods

### Evaluating existing pretrained networks

For input to each network, audio samples were upsampled where required to match the input sample rate of the respective model (Table 1). As samples were shorter than the input window size of BirdNET and Perch, zero-padding was applied to the tail end of each sample. As samples were twice the length of the input window size of VGGish and YAMNet, samples were split into two windows, feature embeddings were calculated for each window and the mean across both taken.

For each of the 16 datasets in ReefSet, each pretrained network was configured with a final fully connected linear layer with corresponding output heads for the classes present in the respective dataset. This final layer was trained for 128 epochs using a batch size of 32, learning rate of 0.001 and categorical cross entropy loss. This was repeated using 4, 8, 16 and 32 training samples per class, with ten repeats using a new random seed for the train-test split and initialization of each. The mean area under the receiver operator curve (AUC-ROC) was calculated across the ten repeats for each of the four training sample counts (Merriënboer, 2023).

### Pretraining with in-domain data

State of the art bioacoustic few-shot learners commonly utilize convolutional neural network architectures (Nolasco et al., 2022). We therefore adapted the pretraining protocol used for Perch, where a CNN classifier is trained and used to extract high quality feature embedding representations for transfer learning (Ghani et al., 2023). During training, datasets were filtered to remove any samples with the primary label 'ambient', in order to eliminate samples that may contain unintentional positive matches for classes in other datasets. For input to the network, samples were upsampled to 32 kHz and log-mel PCEN spectrograms calculated from each (Supp. 1), with repeat padding used to meet the 5s input shape. Samples were shuffled and two augmentations were implemented throughout training: random normalization with a minimum and maximum gain of 0.15 and 0.25, and, MixUp with a mix in probability of 0.75 (Xu et al., 2018). The network was configured with output heads for each primary label (biophony, geophony, anthrophony) and for the 35 secondary labels, less any classes exclusive to held-out data. All training runs were completed for 200K steps.

For the first stage of the experiment a hyperparameter sweep was performed where models were trained on 14 of the 16 datasets with two held-out for validation (Supp. 1). Learning rate, EfficientNet architecture and batch size were probed during the sweep. The core pretraining stage of the experiment was then performed using the optimum hyperparameters from the sweep (Supp. 1. Table S3).

To rigorously test the performance of few-shot transfer learning on unseen datasets we used a Dataset Rotation for Evaluation of Generalization (DREG) approach. During DREG, pretraining was first undertaken using 15 datasets from ReefSet, maintaining one held-out dataset for evaluation. This was repeated in all combinations one by one to produce 16 pretrained models. In the second stage of DREG, evaluation of each model was performed on its respective held-out dataset following the same few-shot transfer learning protocol as described for the existing pretrained networks. Given all data originated from reef habitats, evaluation data could be considered in-domain whilst being out of distribution.

**Pretraining with cross-domain mixing**

The XC Bird dataset was used without modifications to the original pretraining of the Perch model. To integrate this dataset into the training procedure, the model was configured with the same output heads for the ReefSet, alongside additional species, genus, family and order output heads for the XC Bird dataset, with a loss weighting of 0.1 for the latter three compared to standard heads. To integrate the Freesound dataset into pretraining, our only adjustment was to remove all Freesound samples with the label 'bird' (3.38% of the dataset) to mitigate overlap with more taxonomically detailed labels in the XC Bird dataset. The network was then configured as before, alongside an additional set of output heads for Freesound. During pretraining for both the domain mixing strategy experiments, individual datasets were cycled back in once all samples from them had been used once. As with the ReefSet pretraining strategy, a hyperparameter sweep was performed, with an additional parameter for dataset weighting (Supp. 1). All other components remained unchanged, with the DREG approach used for pretraining and evaluation.

**Evaluating SurfPerch on novel bioacoustic domains**

SurfPerch was evaluated in an "off the shelf" manner, with no hyperparameter sweeps or pretraining used to optimize for the novel bioacoustic domains being tested. These novel datasets originated from the bird, bat, frog and marine mammal domains, see Ghani et al. (2023) for further details on the data. As with the other pretraining experiments, a new network was configured with a final classification head to match the target classes for each respective dataset, which was then fine-tuned whilst the rest of the weights were kept frozen. Fine-tuning was conducted for 128 epochs using a batch size of 32, learning rate of 0.001 and categorical cross entropy loss. The fine-tuned networks were then evaluated on held out test sets from their respective dataset. Fine-tuning was performed across multiple counts of training samples per class for each dataset, ranging from 4 to 256, with ten repeats for each count using a new random seed to select the training data.

## Supplementary information

Supplementary 1 contains additional details on the results, methods and ReefSet.

Supplementary 2 contains the SurfPerch model, the ReefSet dataset and a tutorial on using SurfPerch to build a classifier for new data (see code availability): https://doi.org/10.5281/zenodo.11060189

## Data availability

The pretrained SurfPerch model and ReefSet dataset presented in this study are available in the Supplementary 2 Zenodo repository: https://doi.org/10.5281/zenodo.11060189

## Code availability

A detailed tutorial on building classifiers using SurfPerch on new data, with example data provided, is available in the following GitHub repository: https://github.com/BenUCL/surfperch/blob/surfperch/SurfPerch_Demo_with_Calling_in_Our_Corals.ipynb. This tutorial can be accessed and executed download free from a web browser using Google Colaboratory.

The tutorial and associated example data are also available in the Supplementary 2 Zenodo repository: https://doi.org/10.5281/zenodo.11060189. The GitHub repository contains additional code to produce the plots and speed tests.

The Perch codebase used to conduct this study is available from: https://github.com/google-research/perch.

## Acknowledgements

BW was supported by an FSBI PhD Studentship at UCL and ZSL, alongside a PhD Student Researcher placement at Google DeepMind. We wish to acknowledge multiple parties for supporting the collection of data used in this study:

- The United States Virgin Islands datasets were collected in a collaboration between Sound Ocean Science and local partner Corina Marks at Thriving Islands under a USVI Scientific Research Permit: DFW22021X.
- The Mozambique dataset was collected in a collaboration between Sound Ocean Science and local partners Dr Mario LeBrato and Karen Bowles at the Bazaruto Center for Scientific Studies under a Department of Conservation Permit: 04/GDG/ANAC/MTA/2020.

- The Tanzanian dataset was collected in a collaboration between Sound Ocean Science and local partners Dr Mario LeBrato and Karen Bowles at Chumbe at Island Coral Park, Zanzibar, Tanzania under under CHICOP Zanzibar research permits.
- The Thailand dataset was collected under a citizen science program operated by Black Turtle Dive, Thailand.
- The Florida boats dataset was gathered under a Special Activity License (license number: SAL-21-1798-SRP) granted by the Florida Fish and Wildlife Conservation Commission. Funding was provided by a Donald R Nelson Behaviour Research Award to CW by the American Elasmobranch Society.
- The Kenyan dataset was collected in a collaboration between Mars Global and local partners Angus Roberts and Viola Roberts at the Ocean Trust, with permissions from the Lamu County Department of Fisheries.
- Soundscape data from Indonesia were collected as part of the monitoring program for the Mars Coral Reef Restoration Project, in collaboration with Universitas Hasanuddin. We thank Lily Damayanti, Pippa Mansell, David Smith and the Mars Sustainable Solutions team for support with fieldwork logistics. We also thank the Department of Marine Affairs and Fisheries of the Province of South Sulawesi, the Government Offices of the Kabupaten of Pangkep, Pulau Bontosua and Pulau Badi, and the communities of Pulau Bontosua and Pulau Badi for their support. B.W.'s fieldwork in Indonesia was conducted under an Indonesian national research permit issued by BRIN (number 109A/SIP/IV/FR/3/2023), with T.B.R. as the permit's Indonesian researcher/counterpart, and associated ethical approval given by BRIN. We thank Prof J. Jompa and Prof R.A. Rappe at Universitas Hasanuddin for logistical assistance with permit and visa applications.
- Remaining datasets were gathered in a collaboration between Conservation Metrics Inc. and the Cornell Lab of Ornithology with funding support from Oceankind and the Cornell Atkinson Center for Sustainability, Cornell Lab of Ornithology.

## Author information

T.D and B.W conceptualized the study. With supervision from T.D., B.W designed the methodology, implemented the software, conducted the experiments and wrote the manuscript. With supervision from K.J., B.W collected data, annotated data and wrote the manuscript. T.D., B.M., J.H., V.D. and E.T. were responsible for the design and implementation of the software, broader methodology and curation of data. A.F., M.M., J.M. were responsible for curating and annotating data. A.R., A.L., T.R. and C.H. collected data. C.W collected and annotated data.

## Supplementary 1

**Text S1: Inference speed test**

A synthetic audio recording one hour in length sampled at 32 kHz was created. Given models respond differently to batch size and number of threads, a grid search was performed using: batch sizes of 8, 16, 32, 64, or 128, and, 1, 4, 8, 12 or 16 threads. The time to inference over the synthetic audio was recorded and the fastest time across the grid search for each model reported in Table 1. A 13th Gen Intel(R) Core(TM) i9-13900H CPU was used.

**Test S2: PCEN spectrogram settings**

The PCEN spectrograms employed in our experiments used the following parameters:

- Frequency range: 60 - 10'000 hz
- Smoothing Coefficient: 0.1
- Gain: 0.5
- Bias: 2.0
- Root: 2.
- Epsilon: 1e-6
- Spectral PCEN (spcen): False

For SurfPerch, the following modifications were tested after updates to the code base which we found improved performance further:

- Frequency range: 50 - 16'000 hz
- Smoothing Coefficient: 0.145
- Gain: 0.8
- Bias: 10.0
- Root: 4.0

**Text S3: Hyperparameter sweep**

1. For the model pretrained on ReefSet the hyperparameter sweep probed the EfficientNet architecture (b0, b1, b2), learning rate (0.01, 0.001, 0.0001) and batch size (64, 128). During the sweep, fourteen datasets were used for training, with two held-out as validation data. The Kenyan and Bermudan datasets were selected as the validation sets due to their intermediate size and apparent difficulty during the prior experiments using pretrained models (Fig 3; Fig. S1). Resultant networks were evaluated using the mean AUC-ROC from evaluation of the held-out dataset across 4, 8, 16 and 32 training samples per class.
2. For the model pretrained on ReefSet and the XC Bird data all components were the same as those detailed for the ReefSet sweep, with one change which split the sweep into two stages. The first stage once again probed learning rate, alongside the new weighting of bird to reef data (0.1, 0.25, 0.5, 0.75, 0.9). The second stage probed the EfficientNet architecture and batch size as before.

3. For the model pretrained on ReefSet, the XC Bird and the Freesound data the hyperparameter sweep was split into two stages once again, following the ReefSet and XC Bird sweep, with one small adjustment to the first stage; a fixed reef weighting was maintained (0.1) whilst the bird datasets weighting in proportion to all three dataset was probed (0.5, 0.6, 0.7, 0.8), with the remainder assigned to Freesound.

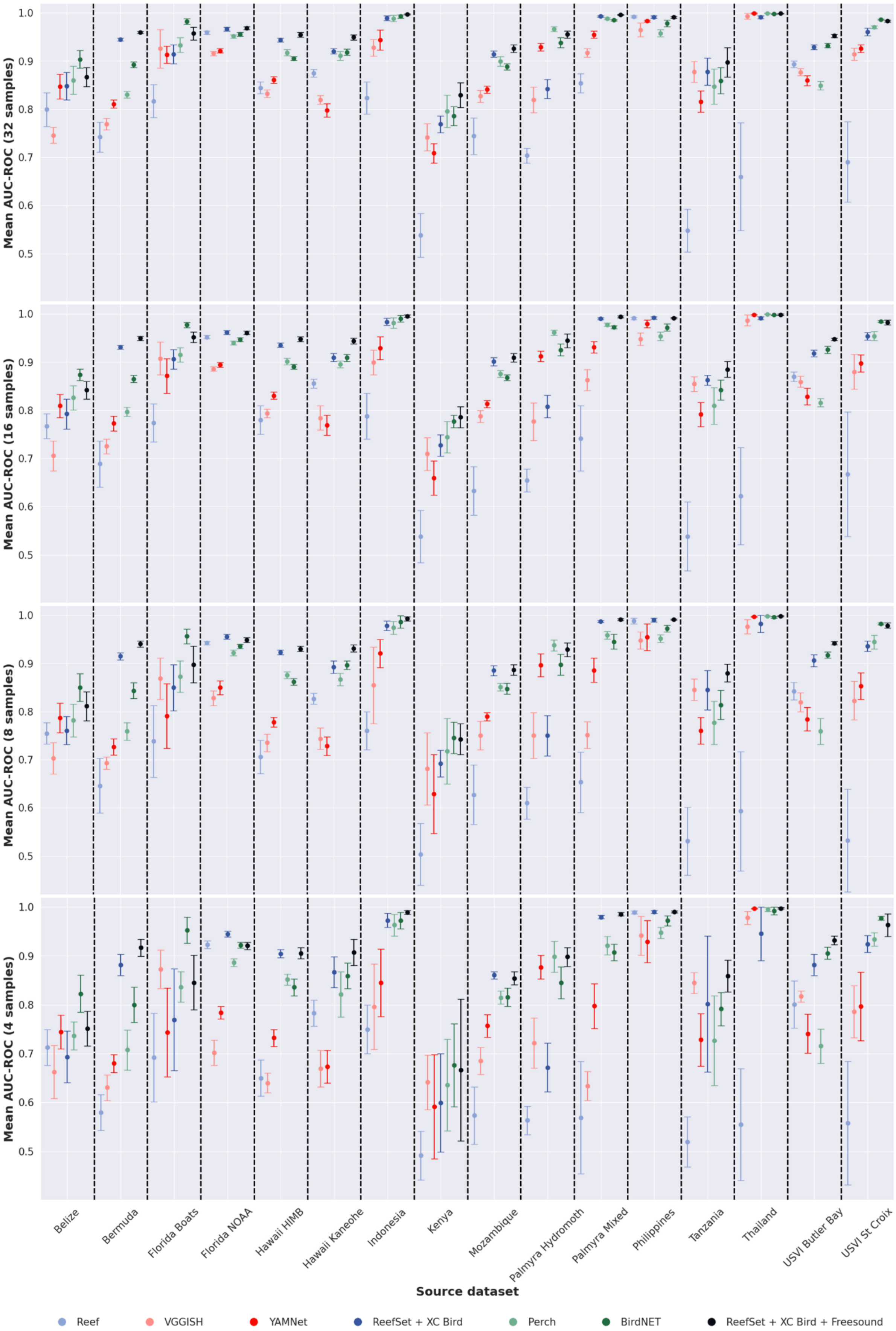

**Fig. S1.** AUC-ROC scores reported by transfer learning evaluation on each dataset within ReefSet using the mean across 10 repeats of 4, 8, 16 and 32 training samples per class. Points represent the mean, lines represent standard deviation of this. Models are ordered by mean performance across all dataset 46 and training sample counts, going from weakest (left) to strongest (right) within each dataset bin. In the 47 legend, existing pretrained networks are indicated by name, whereas our alternative pretraining 48 strategies are indicated by the datasets used during pretraining.

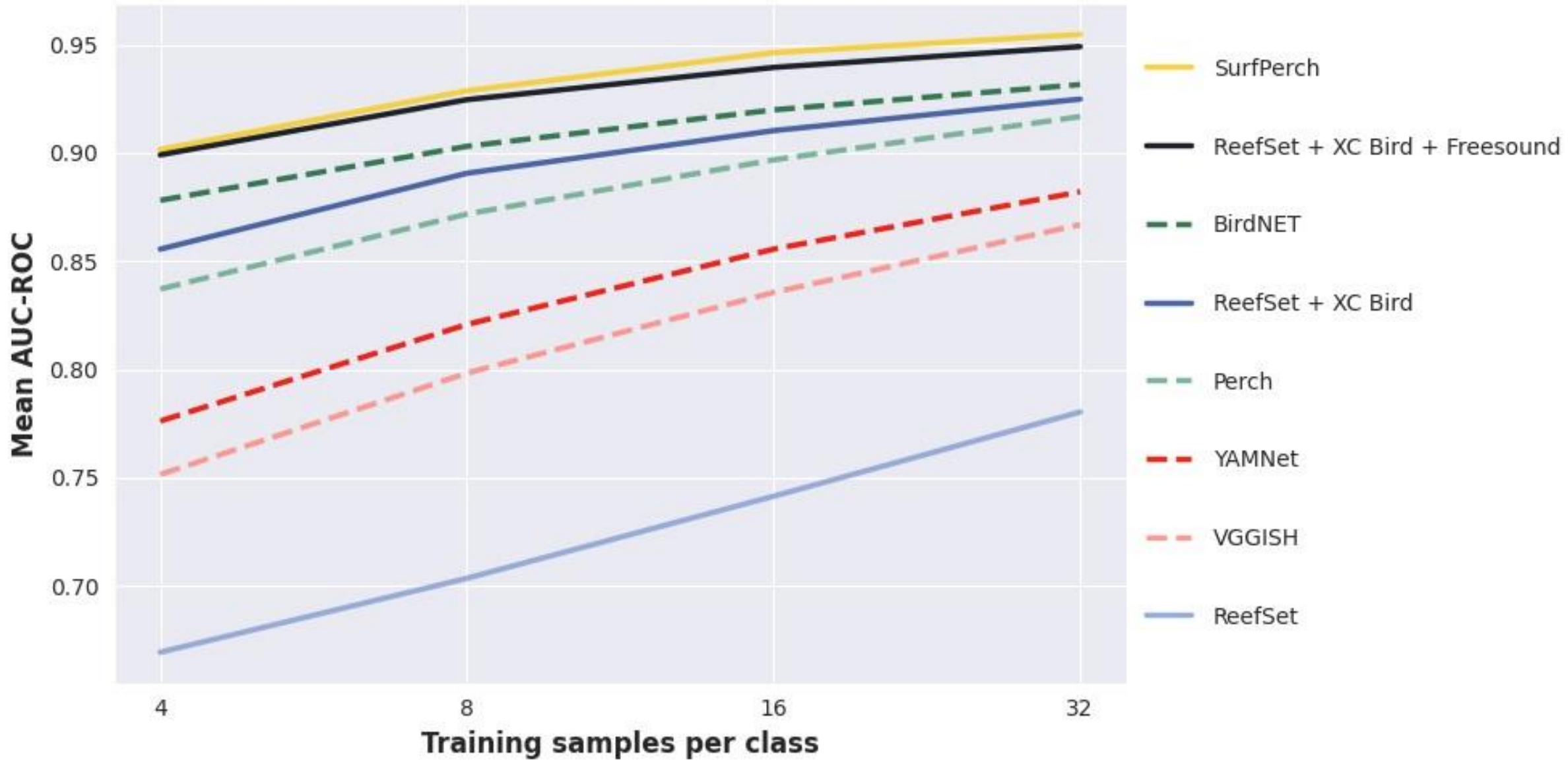

**Fig. S2.** Mean AUC-ROC scores reported by transfer learning evaluation using each model on each of the 52 16 datasets within ReefSet. Dashed lines represent existing pretrained networks, with names indicated 53 in the legend. Thick lines represent our alternative pretraining strategies, with the data used for 54 pretraining indicated in the legend. SurfPerch represents the performance achieved using the modified 55 PCEN spectrogram and training for 1m steps with DREG for evaluation.

**Table S1**. Counts of samples per class label for each dataset within ReefSet.

| Class label | Belize | Bermuda | Florida Boats | Florida | Hawaii HIMB | Hawaii Kaneohe | Indonesia | Kenya | Mozambique | Palmyra hydromoth | Palmyra Mixed | Philippines | Tanzania | Thailand | USVI Butler Bay | USVI St Croix |
|---|---|---|---|---|---|---|---|---|---|---|---|---|---|---|---|---|
| ambient | 0 | 0 | 3528 | 0 | 0 | 0 | 1800 | 563 | 0 | 0 | 0 | 0 | 0 | 0 | 0 | 0 |
| anthrop_boat_engine | 0 | 397 | 1522 | 2117 | 476 | 64 | 0 | 0 | 0 | 0 | 0 | 94 | 0 | 244 | 193 | 0 |
| anthrop_bomb | 0 | 0 | 0 | 0 | 0 | 0 | 203 | 0 | 0 | 0 | 0 | 0 | 0 | 0 | 0 | 0 |
| anthrop_mechanical | 0 | 0 | 0 | 0 | 0 | 0 | 0 | 0 | 0 | 0 | 0 | 0 | 0 | 0 | 619 | 0 |
| bioph | 45 | 346 | 0 | 922 | 250 | 101 | 0 | 258 | 331 | 588 | 0 | 0 | 74 | 0 | 261 | 97 |
| bioph_cascading_saw | 0 | 0 | 0 | 0 | 68 | 195 | 0 | 0 | 0 | 1035 | 8166 | 0 | 0 | 0 | 0 | 0 |
| bioph_chatter | 0 | 0 | 0 | 49 | 0 | 0 | 0 | 0 | 0 | 106 | 0 | 0 | 0 | 0 | 0 | 0 |
| bioph_chorus | 63 | 0 | 0 | 0 | 0 | 0 | 0 | 0 | 0 | 0 | 0 | 0 | 0 | 0 | 0 | 0 |
| bioph_crackle | 203 | 0 | 0 | 0 | 0 | 0 | 0 | 0 | 0 | 0 | 0 | 0 | 0 | 0 | 0 | 0 |
| bioph_croak | 0 | 0 | 0 | 2661 | 0 | 0 | 0 | 0 | 0 | 0 | 0 | 0 | 0 | 0 | 0 | 941 |
| bioph_damselfish | 0 | 0 | 0 | 949 | 0 | 0 | 0 | 0 | 118 | 0 | 245 | 0 | 0 | 0 | 0 | 0 |
| bioph_dolphin | 0 | 0 | 0 | 1748 | 0 | 0 | 0 | 0 | 0 | 0 | 0 | 0 | 0 | 0 | 0 | 0 |
| bioph_double_pulse | 0 | 0 | 0 | 0 | 1099 | 0 | 0 | 0 | 0 | 0 | 0 | 0 | 0 | 0 | 0 | 0 |
| bioph_echinidae | 0 | 0 | 0 | 0 | 0 | 0 | 0 | 0 | 0 | 0 | 1457 | 0 | 0 | 43 | 0 | 0 |
| bioph_epigut | 0 | 326 | 0 | 0 | 0 | 0 | 0 | 0 | 0 | 0 | 0 | 0 | 0 | 0 | 0 | 0 |
| bioph_grazing | 0 | 0 | 0 | 0 | 190 | 0 | 0 | 0 | 0 | 0 | 160 | 0 | 0 | 0 | 0 | 0 |
| bioph_grouper_a | 0 | 102 | 0 | 0 | 0 | 0 | 0 | 0 | 0 | 0 | 0 | 0 | 0 | 0 | 0 | 0 |
| bioph_grouper_groan | 0 | 0 | 0 | 0 | 0 | 0 | 0 | 0 | 140 | 0 | 0 | 0 | 0 | 0 | 0 | 0 |
| bioph_growl | 0 | 0 | 0 | 968 | 0 | 0 | 0 | 0 | 0 | 0 | 1047 | 0 | 0 | 0 | 0 | 0 |
| bioph_holocentrus | 0 | 91 | 0 | 0 | 0 | 0 | 0 | 0 | 0 | 0 | 0 | 0 | 0 | 0 | 0 | 0 |
| bioph_knock | 0 | 0 | 0 | 58 | 1116 | 0 | 0 | 0 | 0 | 0 | 80 | 0 | 0 | 0 | 0 | 87 |
| bioph_knock_croak_a | 0 | 0 | 0 | 0 | 2951 | 0 | 0 | 0 | 0 | 0 | 0 | 0 | 0 | 0 | 0 | 0 |
| bioph_knock_croak_b | 0 | 0 | 0 | 0 | 326 | 0 | 0 | 0 | 0 | 0 | 0 | 0 | 0 | 0 | 0 | 0 |
| bioph_knock_croak_c | 0 | 0 | 0 | 0 | 1384 | 0 | 0 | 0 | 0 | 0 | 0 | 0 | 0 | 0 | 0 | 0 |
| bioph_low_growl | 0 | 0 | 0 | 0 | 542 | 0 | 0 | 0 | 0 | 0 | 0 | 0 | 0 | 0 | 0 | 0 |
| bioph_megnov | 0 | 0 | 0 | 216 | 0 | 0 | 0 | 0 | 271 | 0 | 0 | 0 | 1597 | 0 | 417 | 0 |
| bioph_midshipman | 0 | 0 | 0 | 4660 | 0 | 0 | 0 | 0 | 0 | 0 | 0 | 0 | 0 | 0 | 0 | 0 |
| bioph_mycbon | 0 | 497 | 0 | 233 | 0 | 0 | 0 | 0 | 0 | 0 | 0 | 0 | 0 | 0 | 0 | 0 |
| bioph_pomamb | 0 | 0 | 0 | 0 | 0 | 0 | 0 | 0 | 0 | 0 | 0 | 128 | 0 | 0 | 0 | 0 |
| bioph_pulse | 0 | 0 | 0 | 1108 | 90 | 145 | 0 | 0 | 116 | 0 | 0 | 0 | 0 | 0 | 0 | 0 |
| bioph_rattle | 0 | 0 | 0 | 2508 | 0 | 0 | 0 | 0 | 0 | 0 | 0 | 0 | 0 | 0 | 0 | 0 |
| bioph_rattle_response | 0 | 0 | 0 | 558 | 0 | 0 | 0 | 0 | 0 | 0 | 0 | 0 | 0 | 0 | 0 | 0 |
| bioph_series_a | 0 | 0 | 0 | 174 | 0 | 0 | 0 | 0 | 0 | 0 | 0 | 0 | 0 | 0 | 0 | 0 |
| bioph_series_b | 0 | 0 | 0 | 0 | 0 | 0 | 0 | 0 | 0 | 0 | 65 | 0 | 0 | 0 | 0 | 0 |
| bioph_stridulation | 0 | 0 | 0 | 0 | 83 | 113 | 0 | 0 | 0 | 0 | 184 | 0 | 0 | 0 | 0 | 0 |
| bioph_whup | 0 | 0 | 0 | 0 | 0 | 0 | 0 | 0 | 55 | 0 | 0 | 0 | 0 | 0 | 0 | 0 |
| geoph_waves | 0 | 0 | 0 | 0 | 49 | 0 | 0 | 0 | 0 | 0 | 0 | 0 | 0 | 0 | 0 | 0 |

**Table S2**. Metadata associated with each dataset including the data provider, model of hydrophone used for the dataset and number of unique sites recorded within the dataset.

| Dataset | Data origin | Hydrophones used | Number of sites recorded |
|---|---|---|---|
| Belize | Cornell Lab of Ornithology | SoundTrap 600 | 1 |
| Bermuda | Cornell Lab of Ornithology | SoundTrap 300 | 1 |
| Florida Boats | University of Bristol | Hydromoth | 16 |
| Florida NOAA | NOAA | SoundTrap 500 | 1 |
| Hawaii HIMB | University of Hawai'i at Mānoa, Cornell Lab of Ornithology, Conservation Metrics | SoundTrap 300 | 3 |
| Hawaii Kaneohe | University of Hawai'i at Mānoa, Cornell Lab of Ornithology, Conservation Metrics | SoundTrap 300 | 2 |
| Indonesia | UCL & Mars | Hydromoth | 18 |
| Kenya | UCL & the Ocean Trust | Hydromoth | 5 |
| Mozambique | Sound Ocean Science | SoundTrap 300 | 4 |
| Palmyra Hydromoth | Conservation Metrics, Cornell Lab of Ornithology | Hydromoth | 3 |
| Palmyra Mixed | Conservation Metrics, Cornell Lab of Ornithology | SoundTrap600 & HydroMoth | 10 |
| Philippines | Cornell Lab of Ornithology | SoundTrap 300 | 1 |
| Tanzania | Sound Ocean Science | SoundTrap 300 | 1 |
| Thailand | Sound Ocean Science | Hydromoth | 1 |
| USVI Butler Bay | Sound Ocean Science | SoundTrap 300 | 1 |
| USVI St Croix | Sound Ocean Science | SoundTrap 300 | 1 |

**Table S3.** Results from hyperparameter sweeps detailed in supplementary Text 2. Sweeps are ordered by training data, then stage of sweep where applicable, then mean AUC-ROC score across 4, 8, 16 and 32 training samples. NA indicates the respective hyperparameter was not probed at the matching stage.

| Training data | Stage | Batch size | EfficientNet architecture | Learning rate | Bird weight | AUC-ROC |
|---|---|---|---|---|---|---|
| ReefSet | NA | 128 | b0 | 0.0001 | NA | 0.8 |
| ReefSet | NA | 64 | b0 | 0.0001 | NA | 0.793 |
| ReefSet | NA | 64 | b2 | 0.001 | NA | 0.784 |
| ReefSet | NA | 64 | b0 | 0.001 | NA | 0.78 |
| ReefSet | NA | 128 | b2 | 0.0001 | NA | 0.771 |
| ReefSet | NA | 128 | b1 | 0.0001 | NA | 0.771 |
| ReefSet | NA | 64 | b1 | 0.001 | NA | 0.65 |
| ReefSet | NA | 128 | b2 | 0.01 | NA | 0.749 |
| ReefSet | NA | 128 | b1 | 0.001 | NA | 0.748 |
| ReefSet | NA | 128 | b0 | 0.01 | NA | 0.683 |
| ReefSet | NA | 64 | b2 | 0.0001 | NA | 0.612 |
| ReefSet | NA | 64 | b1 | 0.0001 | NA | 0.611 |
| ReefSet | NA | 128 | b0 | 0.001 | NA | 0.604 |
| ReefSet | NA | 128 | b2 | 0.001 | NA | 0.588 |
| ReefSet | NA | 64 | b1 | 0.01 | NA | 0.57 |
| ReefSet | NA | 64 | b0 | 0.01 | NA | 0.57 |
| ReefSet | NA | 64 | b2 | 0.01 | NA | 0.558 |
| ReefSet | NA | 128 | b1 | 0.01 | NA | 0.556 |
| ReefSet + Bird | 1 | NA | NA | 0.0001 | 0.9 | 0.839 |
| ReefSet + Bird | 1 | NA | NA | 0.001 | 0.9 | 0.824 |
| ReefSet + Bird | 1 | NA | NA | 0.001 | 0.5 | 0.824 |
| ReefSet + Bird | 1 | NA | NA | 0.0001 | 0.75 | 0.821 |
| ReefSet + Bird | 1 | NA | NA | 0.0001 | 0.25 | 0.813 |
| ReefSet + Bird | 1 | NA | NA | 0.0001 | 0.5 | 0.805 |
| ReefSet + Bird | 1 | NA | NA | 0.001 | 0.75 | 0.805 |
| ReefSet + Bird | 1 | NA | NA | 0.001 | 0.1 | 0.803 |
| ReefSet + Bird | 1 | NA | NA | 0.0001 | 0.1 | 0.8 |
| ReefSet + Bird | 1 | NA | NA | 0.01 | 0.75 | 0.781 |
| ReefSet + Bird | 1 | NA | NA | 0.01 | 0.5 | 0.78 |
| ReefSet + Bird | 1 | NA | NA | 0.001 | 0.25 | 0.762 |
| ReefSet + Bird | 1 | NA | NA | 0.01 | 0.25 | 0.754 |
| ReefSet + Bird | 1 | NA | NA | 0.01 | 0.1 | 0.751 |
| ReefSet + Bird | 1 | NA | NA | 0.01 | 0.9 | 0.744 |
| ReefSet + Bird | 2 | 128 | b1 | NA | NA | 0.837 |
| ReefSet + Bird | 2 | 64 | b2 | NA | NA | 0.826 |
| ReefSet + Bird | 2 | 64 | b1 | NA | NA | 0.824 |
| ReefSet + Bird | 2 | 128 | b2 | NA | NA | 0.816 |
| ReefSet + Bird | 2 | 128 | b0 | NA | NA | 0.807 |
| ReefSet + Bird | 2 | 64 | b0 | NA | NA | 0.795 |
| ReefSet + Bird + Freesound | 1 | NA | NA | 0.001 | 0.5 | 0.866 |
| ReefSet + Bird + Freesound | 1 | NA | NA | 0.0001 | 0.5 | 0.858 |
| ReefSet + Bird + Freesound | 1 | NA | NA | 0.001 | 0.6 | 0.849 |
| ReefSet + Bird + Freesound | 1 | NA | NA | 0.001 | 0.7 | 0.847 |
| ReefSet + Bird + Freesound | 1 | NA | NA | 0.0001 | 0.6 | 0.837 |
| ReefSet + Bird + Freesound | 1 | NA | NA | 0.0001 | 0.7 | 0.837 |
| ReefSet + Bird + Freesound | 1 | NA | NA | 0.0001 | 0.8 | 0.834 |
| ReefSet + Bird + Freesound | 1 | NA | NA | 0.001 | 0.8 | 0.83 |
| ReefSet + Bird + Freesound | 1 | NA | NA | 0.01 | 0.5 | 0.828 |
| ReefSet + Bird + Freesound | 1 | NA | NA | 0.01 | 0.7 | 0.807 |
| ReefSet + Bird + Freesound | 1 | NA | NA | 0.01 | 0.6 | 0.802 |
| ReefSet + Bird + Freesound | 1 | NA | NA | 0.01 | 0.8 | 0.798 |
| ReefSet + Bird + Freesound | 2 | 128 | b1 | NA | NA | 0.869 |
| ReefSet + Bird + Freesound | 2 | 128 | b2 | NA | NA | 0.866 |
| ReefSet + Bird + Freesound | 2 | 128 | b0 | NA | NA | 0.841 |
| ReefSet + Bird + Freesound | 2 | 64 | b0 | NA | NA | 0.833 |
| ReefSet + Bird + Freesound | 2 | 64 | b1 | NA | NA | 0.816 |
| ReefSet + Bird + Freesound | 2 | 64 | b2 | NA | NA | 0.814 |